\documentclass[%
 reprint,
 prb,
 aps,
 amsmath,amssymb,
 floatfix,
 superscriptaddress,
 longbibliography,
 nofootinbib
]{revtex4-2}

\usepackage{graphicx}
\usepackage{bm}
\usepackage{dcolumn}
\usepackage{color}
\usepackage{array}
\usepackage{amsmath}
\usepackage{upgreek}

\graphicspath{{figures/}}
\newcommand{\ii}{\mathrm{i}}
\newcommand{\dd}{\mathrm{d}}
\newcommand{\D}{D}

\usepackage{hyperref}
\hypersetup{
 colorlinks=true,
 linkcolor=blue,
 citecolor=blue,
 urlcolor=blue,
}
\usepackage{orcidlink}

\begin{document}

\title{Microscopic calculation of coherence lengths and magnetic penetration depth in multiband superconductors}

\author{Tristan Ryoma Fuchs\,\orcidlink{0009-0008-1109-7449}}
\email{tristan@g.ecc.u-tokyo.ac.jp}
\affiliation{Department of Physics, The University of Tokyo, Tokyo 113-0033, Japan}

\author{Takuya Nomoto\,\orcidlink{0000-0002-4333-6773}}
\affiliation{Department of Physics, Tokyo Metropolitan University, Hachioji 192-0397, Japan}

\author{Hikaru Watanabe\,\orcidlink{0000-0001-7329-9638}}
\affiliation{Graduate School of Engineering, Hokkaido University, Sapporo 060-8628, Japan}

\author{Ryotaro Arita\,\orcidlink{0000-0001-5725-072X}}
\affiliation{Department of Physics, The University of Tokyo, Tokyo 113-0033, Japan}
\affiliation{RIKEN Center for Emergent Matter Science, Wako, Saitama 351-0198, Japan}

\date{\today}

\begin{abstract}
We present an extended Ginzburg--Landau (GL) method for calculating the superconducting coherence length and magnetic penetration depth at temperatures well below the transition temperature $T_{\mathrm c}$. In contrast to conventional GL theory, which expands the free energy in both the order parameters and their gradients, our method applies a perturbative expansion only to the covariant-gradient terms, while retaining the full dependence on the superconducting order parameters. The coefficients of these terms are determined from finite differences of microscopic free energies evaluated at small imposed pair momenta. The method applies to both single-band and multiband superconductors and therefore provides a framework for incorporating more realistic electronic structures. For the models examined here, the extended GL method agrees well with real-space Bogoliubov--de Gennes (BdG) calculations over a wide temperature range, while requiring substantially less computational effort.
\end{abstract}

\maketitle

\section{Introduction}
\label{sec:introduction}

The coherence length $\xi$ and the magnetic penetration depth $\lambda$ are fundamental length scales of a superconductor. $\xi$ sets the length scale over which the superconducting order parameter recovers, whereas $\lambda$ sets the length scale over which the magnetic field decays inside the superconducting state \cite{Ginzburg1950}.  These length scales describe spatial variations in superconductors, including those associated with vortices \cite{Abrikosov1957}.  Their ratio determines the sign of the interface energy between superconducting and normal regions and thus distinguishes type-I from type-II superconductivity.  GL theory, however, is microscopically justified only near the transition temperature $T_{\mathrm c}$, where the order parameter and its spatial gradients are sufficiently small to justify truncating the free-energy expansion \cite{Gorkov1959,Orlova2013}. Consequently, microscopic calculations of $\xi$ and $\lambda$ over a wide temperature range require methods beyond conventional GL theory, which can be computationally demanding.

The need for such calculations is particularly important in multiband superconductors, in which multiple superconducting order parameters are coupled to each other \cite{Suhl1959}. The collective modes of these order parameters can be characterized by multiple coherence lengths \cite{Komendova2011,Komendova2012}, as demonstrated in earlier calculations for a two-orbital negative-$U$ Hubbard model \cite{Litak2012}. When the magnetic penetration depth lies between two coherence lengths, a type-1.5 regime has been proposed \cite{Babaev2005,Silaev2011}. In this regime, vortices can exhibit short-range repulsion and long-range attraction, potentially leading to vortex clustering \cite{Carlstrom2011,Carlstrom2011SemiMeissner,Carlstrom2011ThreeBand}. Experimental signatures of this behavior have been reported in $\mathrm{MgB_2}$ \cite{Nagamatsu2001,An2001,Liu2001,Moshchalkov2009}, although their existence and interpretation remain under debate \cite{Kogan2011,Vagov2016}. A quantitative examination of this scenario requires the determination of all relevant coherence lengths as well as $\lambda$. It is therefore important to calculate these length scales over wide ranges of temperature and model parameters.

Two methods are currently available for calculating these length scales.  First, an approach based on finite-momentum pairing (FMP) uses a superconducting state in which the Cooper pairs have a finite center-of-mass momentum $\bm q$.  It obtains $\xi$ from the $\bm q$ dependence of the order parameter and $\lambda$ from the current response using relations derived from conventional GL theory \cite{Witt2024,Kawamura2026}.  Because their derivation assumes a small order parameter, these relations, and hence this approach, are reliable only when the order parameter is small.

Second, the real-space Bogoliubov--de Gennes (BdG) method instead determines $\xi$ and $\lambda$ directly from the spatial variations of the order parameter and magnetic field around an interface or a vortex \cite{Gygi1991}.  It has also been used to study vortex clustering in a microscopic two-band model \cite{Timoshuk2024}.  The real-space BdG method provides accurate results but requires a self-consistent calculation of the entire vortex or interface on a sufficiently large system size.  A less computationally demanding method that remains applicable well below $T_{\mathrm c}$ is therefore needed.

To address this need, we develop an extended GL method for calculating the superconducting coherence lengths and magnetic penetration depth.  In this method, only the covariant-gradient terms are treated perturbatively, while the full dependence on the superconducting order parameters is retained.  The coefficients of the covariant-gradient terms are extracted from microscopic free energies evaluated at finite $\bm q$ with the order parameters held fixed.  To distinguish the two finite-momentum approaches, we hereafter denote the approach based on conventional GL theory described above as FMP(cGL) and the present extended GL method as FMP(eGL).  FMP(eGL) is applicable at temperatures well below $T_{\mathrm c}$ and to both single-band and multiband superconductors, while requiring substantially less computational effort than real-space BdG calculations.

The paper is organized as follows.  Section~\ref{sec:formulation} presents the multiband BCS model used to calculate the microscopic free energy.  It then derives the single-band and two-band formulations of FMP(eGL) and explains how to extract the corresponding coefficient functions from microscopic free energies at finite $\bm q$.  The section also presents the real-space BdG method used for comparison and the computational scaling and numerical setup.  Section~\ref{sec:results} gives the numerical results for single-band and two-band tight-binding models with isotropic spin-singlet $s$-wave pairing and compares FMP(eGL) with the real-space BdG method, including the two-band case in the type-1.5 regime \cite{Timoshuk2024}.  The results obtained with FMP(eGL) agree well with those obtained with the real-space BdG method over a wide temperature range, whereas FMP(cGL) agrees well with the other two methods only when the order parameter is small.  Section~\ref{sec:conclusion} summarizes the results.  Appendix~\ref{app:fmp} presents the FMP(cGL) formulas derived from conventional GL theory.  Appendix~\ref{app:finite-size} examines finite-size effects on the coherence lengths obtained with the real-space BdG method, and Appendix~\ref{app:nonlocal} examines nonlocal corrections to the magnetic penetration depth obtained with the real-space BdG method.

\section{Formulation}
\label{sec:formulation}

\subsection{FMP(eGL): extended GL method}
\label{subsec:extended-gl-method}

\begin{figure}[!t]
\centering
\includegraphics[width=\columnwidth]{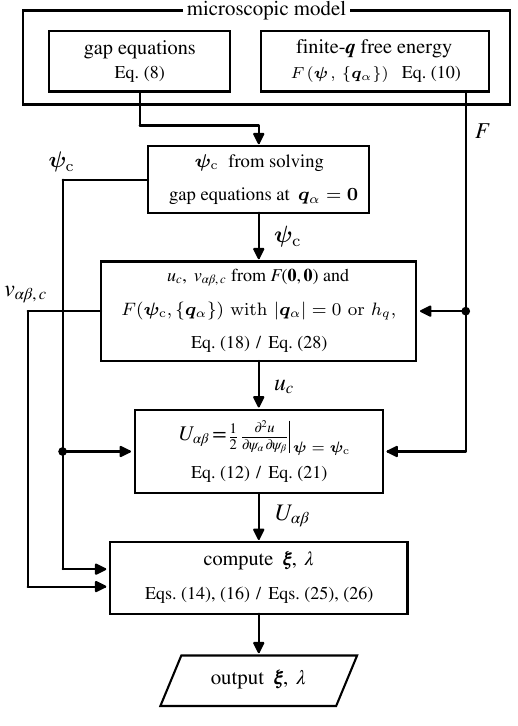}
\caption{Flowchart of FMP(eGL).  Arrows indicate which quantities enter each step.  The order-parameter, uniform-solution, and coherence-length vectors are $\bm\psi=(\psi_1,\ldots,\psi_{N_{\rm b}})^{\mathsf T}$, $\bm\psi_{\mathrm c}=(\psi_{1\mathrm c},\ldots,\psi_{N_{\rm b}\mathrm c})^{\mathsf T}$, and $\bm\xi=(\xi_1,\ldots,\xi_{N_{\rm b}})^{\mathsf T}$, respectively, where $N_{\rm b}$ is the number of bands and $\alpha,\beta=1,\ldots,N_{\rm b}$ are band indices.  The set $\{\bm q_\alpha\}$ denotes the imposed pair momenta.  Equation-reference pairs separated by a slash list the single-band formula first and the two-band formula second.}
\label{fig:flowchart-egl}
\end{figure}

In FMP(eGL), the coherence length ($\xi$) and magnetic penetration depth ($\lambda$) are evaluated from the extended GL free energy, which includes terms to all orders in the order parameter [Eqs.~\eqref{eq:two-band-free-energy} and \eqref{eq:egl-functional}].  Both $\xi$ [Eqs.~\eqref{eq:single-xi} and \eqref{eq:xi-final}] and $\lambda$ [Eqs.~\eqref{eq:single-lambda} and \eqref{eq:lambda-final}] are expressed in terms of the coefficients of the extended GL free energy.  These coefficients are extracted from mean-field calculations based on the microscopic Hamiltonian [Eqs.~\eqref{eq:single-coeff-extract} and \eqref{eq:finiteq-relations}].  In particular, extracting the coefficient of the covariant-gradient term requires evaluating microscopic free energies for Cooper pairs with finite center-of-mass momentum.  The finite-momentum construction follows Ref.~\cite{Witt2024}; in FMP(eGL), however, the order parameters are held fixed during these evaluations, whereas FMP(cGL) determines $\psi_{\bm q}$ self-consistently at each imposed momentum (Appendix~\ref{app:fmp}).  Figure~\ref{fig:flowchart-egl} summarizes the overall procedure.

\subsubsection{Multiband BCS model}
\label{subsec:bcs-meanfield}

We use a multiband spin-singlet BCS Hamiltonian \cite{BCS1957,Suhl1959} as input for FMP(eGL).  Here we derive the corresponding mean-field free energy and gap equations.  The band indices $\alpha$ and $\beta$ run from $1$ to $N_{\rm b}$, where $N_{\rm b}$ is the number of bands.

We consider Cooper pairs in band $\alpha$ with an imposed center-of-mass momentum $\bm q_\alpha$ and denote the full momentum configuration by $\mathcal Q\equiv\{\bm q_\alpha\}_{\alpha=1}^{N_{\rm b}}$.  The momenta of the paired electrons are defined as
\begin{equation}
\bm k_{\alpha\pm}=\pm\bm k+\frac{\bm q_\alpha}{2}
\label{eq:kpm-def}
\end{equation}
The Hamiltonian is
\begin{align}
H=&\sum_{\alpha\bm k\sigma}
\xi_\alpha(\bm k)
 c^\dagger_{\alpha,\bm k,\sigma}
 c_{\alpha,\bm k,\sigma}
\nonumber\\
&-\sum_{\alpha\beta}\sum_{\bm k\bm k'}V_{\alpha\beta}
 c^\dagger_{\alpha,\bm k_{\alpha+},\uparrow}
 c^\dagger_{\alpha,\bm k_{\alpha-},\downarrow}
 c_{\beta,\bm k'_{\beta-},\downarrow}
 c_{\beta,\bm k'_{\beta+},\uparrow} .
\label{eq:bcs-ham}
\end{align}
The one-body kinetic term does not depend on the pair momenta, whereas the pair creation and annihilation operators carry the imposed center-of-mass momenta.  Here $\xi_\alpha(\bm k)$ is the band dispersion measured relative to the chemical potential $\upmu_\alpha$, and $V_{\alpha\beta}$ is the pairing interaction between Cooper pairs in bands $\alpha$ and $\beta$.  The model includes intraband pairing, with pair hopping between bands induced by $V_{\alpha\beta}$ ($\alpha\ne\beta$).  Interband pairing, in which the paired electrons belong to different bands, is excluded.  Momentum conservation for pair hopping between bands $\alpha$ and $\beta$ requires $\bm q_\alpha=\bm q_\beta$.  However, in the two-band formulation of Sec.~\ref{subsec:two-band-egl}, we calculate the free energy at unequal imposed pair momenta to extract the coefficients of the gradient terms independently for each band.

We use the superscript $(\mathcal Q)$ to indicate dependence on $\mathcal Q$.  Applying the standard BCS mean-field approximation to Eq.~\eqref{eq:bcs-ham} gives
\begin{equation}
\begin{split}
H_{\rm MF}=&\sum_{\alpha\bm k}
\begin{pmatrix}
 c^\dagger_{\alpha,\bm k_{\alpha+},\uparrow} &
 c_{\alpha,\bm k_{\alpha-},\downarrow}
\end{pmatrix}
\hat h_{\alpha\bm k}
\begin{pmatrix}
 c_{\alpha,\bm k_{\alpha+},\uparrow}\\
 c^\dagger_{\alpha,\bm k_{\alpha-},\downarrow}
\end{pmatrix}
\\
&+\sum_{\alpha\beta}\Delta_\alpha^{(\mathcal Q)*}
(V^{-1})_{\alpha\beta}\Delta_\beta^{(\mathcal Q)},
\end{split}
\label{eq:two-band-mf-ham}
\end{equation}
The mean-field order parameters in Eq.~\eqref{eq:two-band-mf-ham} are
\begin{equation}
\Delta_\alpha^{(\mathcal Q)}
=\sum_\beta V_{\alpha\beta}
\sum_{\bm k}
\left\langle
c_{\beta,\bm k_{\beta-},\downarrow}
 c_{\beta,\bm k_{\beta+},\uparrow}
\right\rangle .
\label{eq:gapdef}
\end{equation}
A self-consistent solution of the gap equation generally depends on $\mathcal Q$.
The Nambu matrix in Eq.~\eqref{eq:two-band-mf-ham} is
\begin{equation}
\hat h_{\alpha\bm k}=
\begin{pmatrix}
\xi_{\alpha+} & -\Delta_\alpha^{(\mathcal Q)}\\
-\Delta_\alpha^{(\mathcal Q)*} & -\xi_{\alpha-}
\end{pmatrix},
\quad
\xi_{\alpha\pm}=\xi_\alpha\left(\pm\bm k+\frac{\bm q_\alpha}{2}\right).
\label{eq:nambu-matrix}
\end{equation}
Introducing
\begin{equation}
\bar\xi_{\alpha\bm k}^{(\mathcal Q)}=\frac{\xi_{\alpha+}+\xi_{\alpha-}}{2},
\qquad
\delta\xi_{\alpha\bm k}^{(\mathcal Q)}=\frac{\xi_{\alpha+}-\xi_{\alpha-}}{2},
\label{eq:xi-bar-delta}
\end{equation}
we obtain the quasiparticle branches
\begin{equation}
E_{\alpha\bm k}^{\pm}
=\delta\xi_{\alpha\bm k}^{(\mathcal Q)}
\pm
\sqrt{\left[\bar\xi_{\alpha\bm k}^{(\mathcal Q)}\right]^2+|\Delta_\alpha^{(\mathcal Q)}|^2} .
\label{eq:qp-energies}
\end{equation}
The gap equation then reads
\begin{equation}
\Delta_\alpha^{(\mathcal Q)}=\sum_\beta V_{\alpha\beta}
\sum_{\bm k}
\frac{\Delta_\beta^{(\mathcal Q)}}{2E_{\beta\bm k}}
\left[1-f(E_{\beta\bm k}^{+})-f(-E_{\beta\bm k}^{-})\right],
\label{eq:two-band-gap-equation}
\end{equation}
with
\begin{equation}
E_{\beta\bm k}=\sqrt{\left[\bar\xi_{\beta\bm k}^{(\mathcal Q)}\right]^2+|\Delta_\beta^{(\mathcal Q)}|^2}.
\label{eq:Ek-def}
\end{equation}
Here $f(E)=1/[e^{E/(k_\mathrm{B}T)}+1]$ is the Fermi distribution, and $k_\mathrm{B}$ is the Boltzmann constant.
The mean-field free energy is then given by
\begin{equation}
\begin{split}
F^{(\mathcal Q)}=&\sum_{\alpha\beta}\Delta_\alpha^{(\mathcal Q)*}
(V^{-1})_{\alpha\beta}\Delta_\beta^{(\mathcal Q)}
+\sum_{\alpha\bm k}
(\bar\xi_{\alpha\bm k}^{(\mathcal Q)}-E_{\alpha\bm k})\\
&-k_\mathrm{B}T\sum_{\alpha\bm k}
\ln(1+e^{-E_{\alpha\bm k}^{+}/(k_\mathrm{B}T)})\\
&-k_\mathrm{B}T\sum_{\alpha\bm k}
\ln(1+e^{E_{\alpha\bm k}^{-}/(k_\mathrm{B}T)}) .
\end{split}
\label{eq:two-band-free-energy}
\end{equation}
Setting $\Delta_\alpha^{(\mathcal Q)}=0$ and $\bm q_\alpha=\bm 0$ for all $\alpha$ in Eq.~\eqref{eq:two-band-free-energy} gives the normal-state free energy $F_n$.  Equation~\eqref{eq:two-band-gap-equation} determines the equilibrium order parameters and $T_{\mathrm c}$, where $T_{\mathrm c}$ is the temperature above which the nonzero solution at $\bm q_\alpha=\bm 0$ for all $\alpha$ no longer exists.

\subsubsection{Single-band formulation}
\label{subsec:single-formulation}

Throughout this paper, the covariant derivative is defined as $\D=\nabla- \frac{\ii e^*}{\hbar} \bm A$, where $e^*=2e$ is the Cooper-pair charge and $e<0$ is the electron charge. For the single-band case with one order parameter $\psi$, the total free-energy functional of FMP(eGL) is
\begin{equation}
\begin{split}
\mathcal{F}[\psi] =\int \dd\bm r\,
\left[
 F_n+u(\psi)
 +v(\psi)|\D\psi|^2
 +\frac{(\nabla\times\bm A)^2}{2\mu_0}
\right].
\end{split}
\label{eq:single-egl-functional}
\end{equation}
In Eq.~\eqref{eq:single-egl-functional}, $F_n$ is the normal-state free-energy density, and $\mu_0$ is the vacuum permeability. The functions $u(\psi)$ and $v(\psi)$ depend on the order parameter but not on its spatial derivatives.
Contributions of all orders in the order parameter are taken into account, whereas conventional GL theory uses a truncated expansion in powers of the order parameter.  We assume that the order parameter varies slowly on the atomic scale and include $|\D\psi|^2$ as the leading contribution in the covariant-gradient expansion.  Equation~\eqref{eq:single-egl-functional} also omits terms proportional to $[\nabla(|\psi|^2)]^2$, which can contribute when the amplitude varies spatially.  Such terms vanish identically under the uniform phase twists used below, for which the order-parameter amplitude is held fixed.  The coefficients of these amplitude-gradient terms therefore cannot be determined from the present calculations.  The validity of the resulting gradient expansion is examined in Sec.~\ref{subsec:single-result} through comparison with the real-space BdG results.

We now consider small fluctuations around the uniform equilibrium solution, denoted by $\psi_{\mathrm c}$.  The stationarity condition at equilibrium gives
$\left.\partial u/\partial\psi^*\right|_{\psi_{\mathrm c}}=0$.
We choose the equilibrium order parameter to be real and define the coefficient $U$ as
\begin{equation}
U=\frac{1}{2}\left.\frac{\partial^2u}{\partial|\psi|^2}\right|_{\psi_{\mathrm c}}
=\left.\left(\frac{\partial^2u}{\partial\psi^*\partial\psi}
+\frac{\partial^2u}{\partial\psi\partial\psi^*}\right)\right|_{\psi_{\mathrm c}} .
\label{eq:single-U}
\end{equation}
For the uniform solution, we define $u_{\mathrm c}\equiv u(\psi_{\mathrm c})$ and $v_{\mathrm c}\equiv v(\psi_{\mathrm c})$.  For a small real order-parameter fluctuation $\psi=\psi_{\mathrm c}+\delta\psi$ at $\bm A=\bm 0$, setting $\delta \mathcal F = 0$ in FMP(eGL) gives
\begin{equation}
\left[
U-v_{\mathrm c}\nabla^2
\right]\delta\psi=0 .
\label{eq:single-linearized}
\end{equation}
Therefore, derivatives of the coefficient function $v(\psi)$ do not enter Eq.~\eqref{eq:single-linearized}.
With the convention $\nabla^2\delta\psi=\xi^{-2}\delta\psi$, corresponding to $\delta\psi\propto\exp(-x/\xi)$, the single-band coherence length in FMP(eGL) is
\begin{equation}
\xi=
\sqrt{\,
\frac{v_{\mathrm c}}{U}
\,}.
\label{eq:single-xi}
\end{equation}

To derive the magnetic penetration depth, we consider the magnetic response of the uniform state. The gradient term in Eq.~\eqref{eq:single-egl-functional} gives the supercurrent $\bm j=-(2e^{*2}/\hbar^2)v_{\mathrm c}\psi_{\mathrm c}^2\bm A$. Combining this result with Ampère's law gives
\begin{equation}
\left[
\frac{2\mu_0e^{*2}}{\hbar^2}v_{\mathrm c}\psi_{\mathrm c}^2
-\nabla^2
\right]\bm B=0 .
\label{eq:single-magnetic-linearized}
\end{equation}
Comparing this result with the London equation, $\nabla^2\bm B=\lambda^{-2}\bm B$, we obtain the single-band magnetic penetration depth,
\begin{equation}
\lambda=\sqrt{
\frac{\hbar^2}{2\mu_0 e^{*2}v_{\mathrm c}\psi_{\mathrm c}^2}
} .
\label{eq:single-lambda}
\end{equation}

We impose a uniform phase twist on the order parameter, $\psi_{\bm q}(\bm r)=\psi_{\mathrm c}\,e^{\ii\bm q\cdot\bm r}$, keeping its magnitude fixed at $\psi_{\mathrm c}$ while varying only the imposed momentum $q=|\bm q|$.  For sufficiently small $q$,
\begin{equation}
F(\psi_{\mathrm c},q)=F_n+u_{\mathrm c}
+v_{\mathrm c}q^2\psi_{\mathrm c}^2.
\label{eq:single-finiteq}
\end{equation}
Here $F$ denotes the free-energy density.  For the single-band case, the order parameter $\psi$ used in the GL theory is taken to be the mean-field order parameter $\Delta_1^{(\mathcal Q)}$.  We use single-band BCS theory to calculate $F(\psi,q)$ and solve the gap equation at $\bm q=\bm 0$ for $\psi_{\mathrm c}$.  The coefficients $u_{\mathrm c}$ and $v_{\mathrm c}$ are extracted from microscopic free energies at finite $\bm q$.  The numerical evaluation of $U$ is described in Sec.~\ref{subsec:cost-scaling}.

We choose a small momentum step, $q=h_q$.  The choice of $h_q$ is described in Sec.~\ref{subsec:cost-scaling}.  For fixed $\psi_{\mathrm c}$, the three configurations that determine $F_n$, $u_{\mathrm c}$, and $v_{\mathrm c}$ are
\begin{subequations}
\label{eq:single-coeff-extract}
\begin{align}
F(0,0)=&F_n,\\
F(\psi_{\mathrm c},0)=&F_n+u_{\mathrm c},\\
F(\psi_{\mathrm c},h_q)=&F_n+u_{\mathrm c}+v_{\mathrm c}\,h_q^2\psi_{\mathrm c}^2.
\end{align}
\end{subequations}
These formulas are used for the single-band results in Sec.~\ref{subsec:single-result}.

\subsubsection{Two-band formulation}
\label{subsec:two-band-egl}
For two-band systems, the free-energy functional of FMP(eGL) is
\begin{equation}
\begin{split}
\mathcal{F}[\psi_1, \psi_2]=&\int\dd\bm r\Big[
 F_n+u(\psi_1,\psi_2)\\
&+v_{11}(\psi_1,\psi_2)|\D\psi_1|^2
+v_{22}(\psi_1,\psi_2)|\D\psi_2|^2\\
&+v_{12}(\psi_1,\psi_2)
\D\psi_1\cdot(\D\psi_2)^*\\
&+v_{12}^*(\psi_1,\psi_2)
(\D\psi_1)^*\cdot\D\psi_2
+\frac{(\nabla\times\bm A)^2}{2\mu_0}
\Big].
\end{split}
\label{eq:egl-functional}
\end{equation}
The conventional two-band GL functional, with constant coefficients, has been derived microscopically near $T_{\mathrm c}$ \cite{Tilley1964,Zhitomirsky2004}.  Here, in contrast, the functions $u$, $v_{11}$, $v_{22}$, and $v_{12}$ depend on the order parameters but not on their spatial derivatives.  The coefficients $v_{11}$ and $v_{22}$ are real, while $v_{12}$ is in general complex.  Equation~\eqref{eq:egl-functional} likewise omits terms proportional to $[\nabla(|\psi_\alpha|^2)]^2$ and the interband cross term $\nabla(|\psi_1|^2)\cdot\nabla(|\psi_2|^2)$, which can contribute when the amplitudes vary spatially.  The validity of the resulting gradient expansion is likewise examined in Sec.~\ref{subsec:two-result} through comparison with the real-space BdG results.  As in the single-band case, $\psi_{1\mathrm c}$ and $\psi_{2\mathrm c}$ denote the uniform solution.  The stationary condition also holds, $\left.\partial u/\partial\psi_\alpha^*\right|_{\psi_{1\mathrm c},\psi_{2\mathrm c}}=0$.

For two-band systems, the derivation follows the same steps as in the single-band case.  For small real order-parameter fluctuations,
\begin{equation}
\psi_\alpha(\bm r)=\psi_{\alpha\mathrm c}+\delta\psi_\alpha(\bm r),
\label{eq:amplitude-fluctuation}
\end{equation}
for $\alpha,\beta=1,2$, we define the two-band counterpart of Eq.~\eqref{eq:single-U} as
\begin{equation}
U_{\alpha \beta}=\frac{1}{2}\left.\frac{\partial^2u}{\partial\psi_{\alpha}\,\partial\psi_{\beta}}\right|_{\psi_{1\mathrm c},\psi_{2\mathrm c}} .
\label{eq:U-def}
\end{equation}
At the uniform solution, we define $u_{\mathrm c}\equiv u(\psi_{1\mathrm c},\psi_{2\mathrm c})$ and $v_{\alpha\beta,\mathrm c}\equiv v_{\alpha\beta}(\psi_{1\mathrm c},\psi_{2\mathrm c})$ for $\alpha,\beta=1,2$.
For real order parameters, $u$ is real and $U_{21}=U_{12}$.  We therefore use a symmetric $U$ matrix below.
As in the single-band case, derivatives of $v_{11}$, $v_{22}$, and $v_{12}$ do not enter at this order.  Setting $\bm A=\bm 0$ and varying Eq.~\eqref{eq:egl-functional}, we obtain the linearized equations
\begin{equation}
\left[
\begin{pmatrix}
U_{11} & U_{12}\\
U_{12} & U_{22}
\end{pmatrix}
-
\begin{pmatrix}
v_{11,\mathrm c} & \operatorname{Re}[v_{12,\mathrm c}]\\
\operatorname{Re}[v_{12,\mathrm c}] & v_{22,\mathrm c}
\end{pmatrix}
\nabla^2
\right]
\binom{\delta\psi_1}{\delta\psi_2}=0 .
\label{eq:lin-eq}
\end{equation}
Equation~\eqref{eq:lin-eq} gives
\begin{equation}
\det
\begin{pmatrix}
U_{11}-v_{11,\mathrm c}\xi^{-2} & U_{12}-\operatorname{Re}[v_{12,\mathrm c}]\,\xi^{-2}\\
U_{12}-\operatorname{Re}[v_{12,\mathrm c}]\,\xi^{-2} & U_{22}-v_{22,\mathrm c}\xi^{-2}
\end{pmatrix}=0 .
\label{eq:detxi}
\end{equation}
Defining
\begin{subequations}
\label{eq:ABCdef}
\begin{align}
\mathcal A&=v_{11,\mathrm c}v_{22,\mathrm c}-\operatorname{Re}[v_{12,\mathrm c}]^2,\\
\mathcal B&=-U_{11}v_{22,\mathrm c}-U_{22}v_{11,\mathrm c}
+2U_{12}\operatorname{Re}[v_{12,\mathrm c}],\\
\mathcal C&=U_{11}U_{22}-U_{12}^2,
\end{align}
\end{subequations}
we solve Eq.~\eqref{eq:detxi} as a quadratic equation for $\xi^{-2}$ and obtain two solutions,
\begin{equation}
\xi_{1,2}=\sqrt{
\frac{2\mathcal A}{-\mathcal B\pm\sqrt{\mathcal B^2-4\mathcal A\mathcal C}}
} .
\label{eq:xi-final}
\end{equation}
Two coherence lengths were also obtained within conventional GL theory derived from a two-orbital negative-$U$ Hubbard model \cite{Litak2012}.
The two coherence lengths obtained from Eq.~\eqref{eq:detxi} are real and positive if $U_{11}>0$, $U_{11}U_{22}>U_{12}^{2}$, $v_{11,\mathrm c}>0$, and $v_{11,\mathrm c}v_{22,\mathrm c}>\operatorname{Re}[v_{12,\mathrm c}]^{2}$.

As in the single-band case, the magnetic penetration depth is obtained from the London equation:
\begin{equation}
\begin{split}
\lambda={}&\frac{\hbar}{\sqrt{2\mu_0}\,|e^*|}
\big[v_{11,\mathrm c}\psi_{1\mathrm c}^2+v_{22,\mathrm c}\psi_{2\mathrm c}^2\\
&\quad+2\operatorname{Re}[v_{12,\mathrm c}]\,\psi_{1\mathrm c}\psi_{2\mathrm c}\big]^{-1/2} .
\end{split}
\label{eq:lambda-final}
\end{equation}
Compared with the single-band case, Eq.~\eqref{eq:lambda-final} contains an additional contribution from the mixed gradient term between $\psi_1$ and $\psi_2$.  We therefore introduce separate imposed pair momenta $\bm q_1$ and $\bm q_2$ and denote the free-energy density by $F(\psi_1,\psi_2,\bm q_1,\bm q_2)$.  For fixed order parameters $\psi_{1\mathrm c}$ and $\psi_{2\mathrm c}$ and sufficiently small $q_1=|\bm q_1|$ and $q_2=|\bm q_2|$, the free energy can be expanded as
\begin{equation}
\begin{split}
F(\psi_{1\mathrm c},\psi_{2\mathrm c},\bm q_1, \bm q_2)\approx &F_n+u_{\mathrm c}
+v_{11,\mathrm c}q_1^2\psi_{1\mathrm c}^2\\
&+v_{22,\mathrm c}q_2^2\psi_{2\mathrm c}^2\\
&+2\operatorname{Re}[v_{12,\mathrm c}]\,\psi_{1\mathrm c}\psi_{2\mathrm c}q_1q_2 .
\end{split}
\label{eq:finiteq-expansion}
\end{equation}
The order parameters $\psi_\alpha$ used in the GL theory are taken to be the mean-field order parameters $\Delta_\alpha^{(\mathcal Q)}$.  We use the multiband BCS model of Sec.~\ref{subsec:bcs-meanfield} to calculate $F(\psi_1,\psi_2,\bm q_1,\bm q_2)$ and solve the gap equation at $\bm q_1=\bm q_2=\bm 0$ for $\psi_{1\mathrm c}$ and $\psi_{2\mathrm c}$.  To extract $u_{\mathrm c}$, $v_{11,\mathrm c}$, $v_{22,\mathrm c}$, and $\operatorname{Re}[v_{12,\mathrm c}]$, we keep the order parameters fixed at $\psi_{1\mathrm c}$ and $\psi_{2\mathrm c}$ and vary only the imposed pair momenta $(\bm q_1,\bm q_2)$.  The numerical evaluation of the $U_{\alpha \beta}$ is described in Sec.~\ref{subsec:cost-scaling}.

We take each imposed momentum to be either $0$ or a small step $h_q$.  For fixed nonzero order parameters $(\psi_{1\mathrm c},\psi_{2\mathrm c})$, the five configurations used to determine $u_{\mathrm c}$, $v_{11,\mathrm c}$, $v_{22,\mathrm c}$, and $\operatorname{Re}[v_{12,\mathrm c}]$ are
\begin{subequations}
\label{eq:finiteq-relations}
\begin{align}
F(0,0,0,0)=&F_n,\\
F(\psi_{1\mathrm c},\psi_{2\mathrm c},0,0)=&F_n+u_{\mathrm c},\\
F(\psi_{1\mathrm c},\psi_{2\mathrm c},h_q,0)=&F_n+u_{\mathrm c}+v_{11,\mathrm c}h_q^2\psi_{1\mathrm c}^2,\\
F(\psi_{1\mathrm c},\psi_{2\mathrm c},0,h_q)=&F_n+u_{\mathrm c}+v_{22,\mathrm c}h_q^2\psi_{2\mathrm c}^2,\\
F(\psi_{1\mathrm c},\psi_{2\mathrm c},h_q,h_q)=&F_n+u_{\mathrm c}+v_{11,\mathrm c}h_q^2\psi_{1\mathrm c}^2\nonumber\\
&+v_{22,\mathrm c}h_q^2\psi_{2\mathrm c}^2\nonumber\\
&+2\operatorname{Re}[v_{12,\mathrm c}]\,\psi_{1\mathrm c}\psi_{2\mathrm c}h_q^2.
\end{align}
\end{subequations}

These are the two-band formulas used in Sec.~\ref{subsec:two-result}.  In both single-band and two-band systems, the gradient expansion is valid when the order parameters vary slowly on the lattice scale.

\subsection{Real-space BdG method}
\label{subsec:realspace}

\begin{figure}[!t]
\centering
\includegraphics[width=\columnwidth]{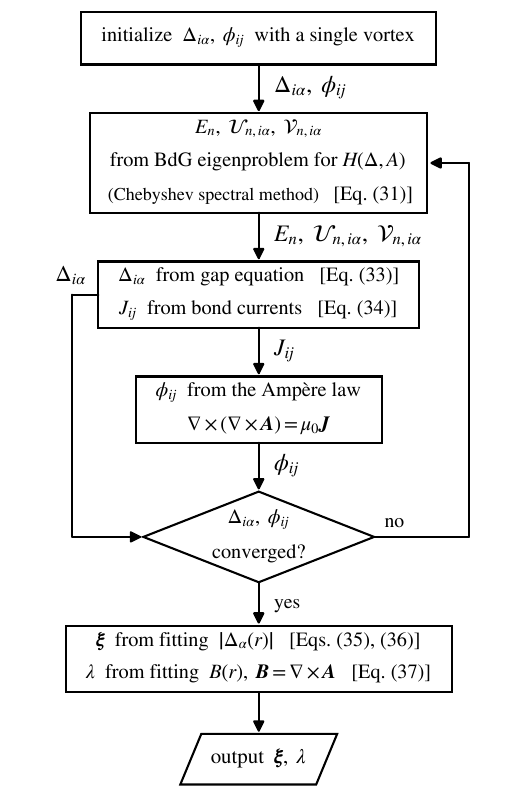}
\caption{Flowchart of the real-space BdG method.  The loop indicates the self-consistent iteration of the order parameter $\Delta_{i\alpha}$ and $\phi_{ij}$.}
\label{fig:flowchart-bdg}
\end{figure}

The real-space BdG method solves the BdG equations on a large real-space lattice \cite{deGennes1966}.  Its results are used for comparison with those obtained with FMP(eGL) in Sec.~\ref{sec:results}.  Figure~\ref{fig:flowchart-bdg} summarizes the procedure of the real-space BdG method.  The calculation is initialized with a single vortex at the center ($\bm r=\bm 0$).  We solve the coupled gap and current equations self-consistently until both $\Delta_{i\alpha}$ and $\phi_{ij}$ converge.  After convergence, we fit the order parameters $|\Delta_\alpha(r)|$ and the magnetic field $B(r)$ to obtain the coherence lengths and magnetic penetration depth.  The equations used in this procedure are presented below.

The real-space BdG method can be applied to both single-band and two-band systems.  The mean-field Hamiltonian with the Peierls phases is

\begin{equation}
\begin{split}
H=&-\sum_{\langle ij\rangle,\sigma,\alpha}
 t_{ij}^{(\alpha)}\exp\!\left[\frac{\ii e}{\hbar} \phi_{ij}\right]
 c_{i\alpha\sigma}^\dagger c_{j\alpha\sigma}\\
&-\sum_{i,\sigma,\alpha}\upmu_\alpha
 c_{i\alpha\sigma}^\dagger c_{i\alpha\sigma}\\
&-\sum_{i,\alpha}
\left(\Delta_{i\alpha}
 c_{i\alpha\uparrow}^\dagger c_{i\alpha\downarrow}^\dagger
+{\mathrm{H.c.}}\right) .
\end{split}
\label{eq:bdg-hij}
\end{equation}
Here $t_{ij}^{(\alpha)}$ is the hopping amplitude in band $\alpha$.  The quantity $\phi_{ij}$ is defined by
\begin{equation}
\phi_{ij}=\int_j^i \bm A\cdot \dd\bm \ell .
\label{eq:bdg-peierls}
\end{equation}
The matrix $H_{ij}^{(\alpha)}$ is then defined as
\begin{equation}
H_{ij}^{(\alpha)}=-t_{ij}^{(\alpha)}\exp\!\left[\frac{\ii e}{\hbar} \phi_{ij}\right]-\upmu_\alpha\delta_{ij}.
\label{eq:bdg-single-particle}
\end{equation}
The self-consistent order parameters are
\begin{equation}
\Delta_{i\alpha}=\sum_\beta V_{\alpha\beta}
\langle c_{i\beta\downarrow}c_{i\beta\uparrow}\rangle,
\label{eq:bdg-gap-self}
\end{equation}
with the same definition as in Eq.~\eqref{eq:gapdef}.  For eigenvalue $E_n$ and eigenvector components $\mathcal{U}_{n,i\alpha}$ and $\mathcal{V}_{n,i\alpha}$, the eigenvalue problem of the real-space BdG method is
\begin{equation}
\sum_j
\begin{pmatrix}
H_{ij}^{(\alpha)} & -\Delta_{i\alpha}\delta_{ij}\\
-\Delta_{i\alpha}^*\delta_{ij} & -H_{ij}^{(\alpha)*}
\end{pmatrix}
\begin{pmatrix}
\mathcal{U}_{n,j\alpha}\\ \mathcal{V}_{n,j\alpha}
\end{pmatrix}
=E_n
\begin{pmatrix}
\mathcal{U}_{n,i\alpha}\\ \mathcal{V}_{n,i\alpha}
\end{pmatrix} .
\label{eq:bdg-equation}
\end{equation}
The bond current is
\begin{equation}
J_{ij}=-\frac{2e}{\hbar}\sum_{\alpha\sigma}
{\mathrm{Im}}\left[
t_{ij}^{(\alpha)}\exp\!\left[\frac{\ii e}{\hbar} \phi_{ij}\right]
\langle c_{i\alpha\sigma}^\dagger c_{j\alpha\sigma}\rangle
\right].
\label{eq:bdg-bond-current}
\end{equation}
$\phi_{ij}$ is calculated from $J_{ij}$ using Amp\`ere's law, $\nabla\times(\nabla\times\bm A)=\mu_0\,\bm J$, and the magnetic field is defined by $\bm B=\nabla\times\bm A$.  The total energy is the sum of the electronic free energy associated with Eq.~\eqref{eq:bdg-hij} and the magnetic energy $\frac{1}{2\mu_0}\sum \bm B^2$.  We update $\phi_{ij}$ using the stationary condition for the total energy.

In the fitting functions below, $r=|\bm r|$ is the distance from the vortex center and $B$ denotes the $z$ component of $\bm B$.  For a vortex in the single-band model, we fit $|\Delta(r)|$ to
\begin{equation}
|\Delta(r)|=|\Delta^u|-C_\Delta K_0(r/\xi),
\label{eq:fit-single}
\end{equation}
where $|\Delta^u|$ is the bulk value of the order parameter far from the vortex core and $K_0$ is the modified Bessel function of the second kind.  For a two-band system, we use the form~\cite{Timoshuk2024}
\begin{subequations}
\label{eq:fit-two}
\begin{equation}
|\Delta_1(r)|=|\Delta_1^u|
-C_1\cos\Theta\,K_0(r/\xi_1)
+C_2\sin\Theta\,K_0(r/\xi_2),
\end{equation}
\begin{equation}
|\Delta_2(r)|=|\Delta_2^u|
-C_1\sin\Theta\,K_0(r/\xi_1)
-C_2\cos\Theta\,K_0(r/\xi_2).
\end{equation}
\end{subequations}
In Eqs.~\eqref{eq:fit-single}--\eqref{eq:fit-B}, the coefficients $C_\Delta$, $C_1$, $C_2$, $C_B$, the mixing angle $\Theta$, $|\Delta^u|$ and $|\Delta_\alpha^u|$ are the fitting parameters.  The factors $\cos\Theta$ and $\sin\Theta$ give the relative weights of the two $K_0$ terms of Eq.~\eqref{eq:fit-two} in the two order parameters.  We fit the magnetic field to the solution of the London equation,
\begin{equation}
B(r)=C_B\,K_0(r/\lambda).
\label{eq:fit-B}
\end{equation}
Nonlocal corrections to this fitting procedure are discussed in Appendix~\ref{app:nonlocal}.  This form applies to both the single-band and two-band cases.

\subsection{Computational scaling and numerical setup}
\label{subsec:cost-scaling}

\begin{table}
\caption{Computational cost of the three methods for the single-band model.  Here $N_k$, $N_q$, and $N_{\rm Ch}$ are the numbers of $k$ points, imposed momenta ($N_q=3$ here), and Chebyshev moments, $N_{\rm sc}$ is the number of iterations of the self-consistent gap equation, $N_{\rm it}$ is the number of BdG self-consistency iterations, and $L$ is the linear lattice size.}
\label{tab:cost}
\centering
\begingroup
\renewcommand{\arraystretch}{1.12}
\begin{tabular*}{0.82\columnwidth}{@{\extracolsep{\fill}}lc}
\hline\hline
Method & Cost per $T$ point \\
\hline
FMP(cGL) & $O(N_qN_{\rm sc}N_k)$ \\
FMP(eGL) & $O(N_{\rm sc}N_k)$ \\
real-space BdG method & $O(N_{\rm it}N_{\rm Ch}L^4)$ \\
\hline\hline
\end{tabular*}
\endgroup
\end{table}

We use the following numerical settings for the calculations of Sec.~\ref{sec:results}.  Here $N_k$ is the number of $\bm k$ points, $N_q$ is the number of imposed momenta, and $N_{\rm sc}$ is the number of iterations of the self-consistent gap equation.  For FMP(eGL), we evaluate microscopic free energies at finite $\bm q$ on $N_k=500\times500$ (single-band model) and $N_k=1500\times1500$ (two-band model) momentum meshes.  We solve the gap equation self-consistently until the absolute difference in $\psi_\alpha$ between successive iterations is below $10^{-8}$.  We calculate the derivatives of $u$ using central finite differences about the self-consistent solution $\psi_{\alpha\mathrm c}$ with step $h_\psi=10^{-3}$.  The required free-energy evaluations at fixed order parameters add negligible computational cost.  The momentum step in Eqs.~\eqref{eq:single-coeff-extract} and \eqref{eq:finiteq-relations} is $h_q=10^{-4}$.

We use $L\times L$ lattices with $L=64$ for all calculations with the real-space BdG method.  We solve the gap equation and the current equation~\eqref{eq:bdg-bond-current} self-consistently until the average absolute differences between successive iterations are below $10^{-8}$ for $\Delta_{i\alpha}$ and $10^{-6}$ for $\phi_{ij}$.

In all fits, the vortex center is not taken into account because Eqs.~\eqref{eq:fit-single}--\eqref{eq:fit-B} do not apply at the vortex core; positions within $\simeq2\,\xi_\alpha$ around the center are excluded from fits to $|\Delta_\alpha(r)|$, where $\xi_\alpha$ is the fitted coherence length of more dominant band.  Likewise, data within $\simeq2\,\lambda$ are excluded from fits to $B(r)$.  The outer edge of each fitting range is chosen where the data no longer follow the corresponding equation (see Appendix~\ref{app:nonlocal}).  Appendix~\ref{app:finite-size} verifies that the system size is sufficient.

We avoid full diagonalization, which would scale as $O(L^6)$ per self-consistency step in two dimensions, by using the Chebyshev spectral method \cite{Covaci2010,Nagai2012}.  We use $N_{\rm Ch}=400$ Chebyshev moments, and convergence typically requires $(1\text{--}4)\times10^3$ self-consistency iterations.

Table~\ref{tab:cost} summarizes the leading computational cost of each method.  For fixed $N_q$, FMP(cGL), described in Appendix~\ref{app:fmp}, and FMP(eGL) have the same asymptotic dependence on $N_{\rm sc}$ and $N_k$.  The real-space BdG method requires solving the equations on an $L\times L$ real-space lattice and scales as $L^4$.

\section{Results}
\label{sec:results}

Throughout this section, we set the lattice constant to $a=1$ and the nearest-neighbor hopping amplitude to $t=1$.  Following the rescaling introduced in Appendix A of Ref.~\cite{Benfenati2023}, we define the effective two-dimensional vacuum permeability as $\mu_0^{2\mathrm D}=\mu_0/L_z$, where $L_z$ is the effective thickness of the two-dimensional system.  The dimensionless charge parameter $e'$ and the rescaled magnetic field $B'$ are then defined as $e'=(ea/\hbar)\sqrt{\mu_0^{2\mathrm D}t}$ and $B'=Ba/\sqrt{\mu_0^{2\mathrm D}t}$, respectively.  All numerical results below are expressed in terms of the rescaled charge $e'$ and magnetic field $B'$.

\subsection{Single-band model}
\label{subsec:single-result}

\begin{figure}[t]
\centering
\includegraphics[width=0.9\linewidth]{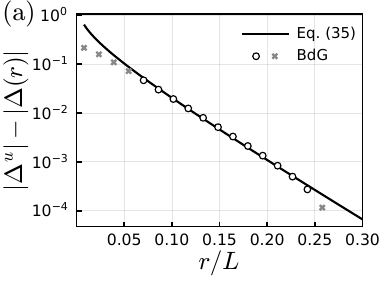}

\medskip

\includegraphics[width=0.9\linewidth]{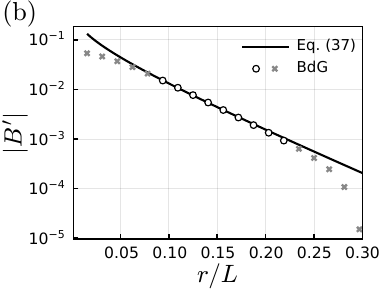}
\caption{Fits to (a) $|\Delta(r)|$ and (b) $B'(r)$ obtained with the real-space BdG method for the single-band model of Sec.~\ref{subsec:single-result} ($V=2.0$, $L=64$, $T/T_{\mathrm c}=0.8$).  The radial coordinate is normalized by $L=64$.  Open circles indicate the data included in the fits, and gray crosses indicate the excluded data.}
\label{fig:bdg-fit-single}
\end{figure}

\begin{figure}[!t]
\centering
\includegraphics[width=0.9\linewidth]{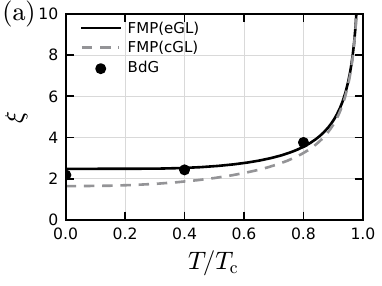}

\medskip

\includegraphics[width=0.9\linewidth]{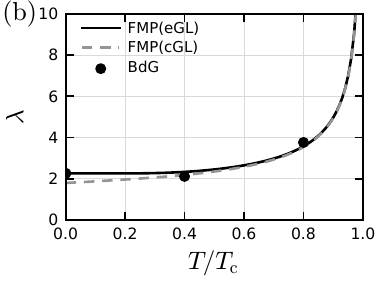}
\caption{Results for the single-band model with $V=2.0$.  (a) Coherence length and (b) magnetic penetration depth as functions of $T/T_{\mathrm c}$.  Solid curves show the results of FMP(eGL), dashed curves show the results of FMP(cGL), and symbols show selected values obtained with the real-space BdG method.}
\label{fig:single-results}
\end{figure}

We first apply FMP(eGL) to a single-band BCS model on a two-dimensional square lattice with pairing strength $V=2.0$.  All three methods use the dispersion
\begin{equation}
\xi(\bm k)=-2t[\cos(k_xa)+\cos(k_ya)]-\upmu,
\label{eq:tight-binding-dispersion}
\end{equation}
with $\upmu=0.0$ and $|e'|=0.5$.  Representative fits to $|\Delta(r)|$ and $B'(r)$ obtained with the real-space BdG method [Eqs.~\eqref{eq:fit-single} and \eqref{eq:fit-B}] at $T/T_{\mathrm c}=0.8$ are shown in Fig.~\ref{fig:bdg-fit-single}.  Figure~\ref{fig:single-results} compares $\xi$ and $\lambda$ obtained with the three methods as functions of temperature.

The results obtained with FMP(eGL) and the real-space BdG method agree well at all temperatures studied for $\xi$ and over the range shown for $\lambda$.  Even at the lowest temperatures, $\xi$ is only a few lattice constants.  This agreement indicates that the gradient expansion remains valid at this length scale.  At low $T/T_{\mathrm c}$, where $\kappa=\lambda/\xi$ becomes small, the magnetic response becomes nonlocal.  Appendix~\ref{app:nonlocal} verifies at $T/T_{\mathrm c}=0.4$ that the fit based on Eq.~\eqref{eq:fit-B} using only a few data points still gives a reasonable estimate of the magnetic penetration depth.  FMP(cGL), described in Appendix~\ref{app:fmp}, agrees well with the other two methods only when the order parameter is small for both length scales.

\subsection{Two-band model}
\label{subsec:two-result}

We next consider a two-band BCS model with pairing strengths
\[
V_{11}=2.8,\qquad V_{22}=2.2,\qquad V_{12}=V_{21}=0.01.
\]
For each band $\alpha=1,2$, the dispersion has the same square-lattice form,
\begin{equation}
\xi_\alpha(\bm k)=-2t_\alpha[\cos(k_xa)+\cos(k_ya)]-\upmu_\alpha,
\label{eq:two-band-dispersion}
\end{equation}
with $t_1=t_2=t=1$, $\upmu_\alpha=0.0$, and charge parameter $|e'|=0.6$, matching the parameters of Ref.~\cite{Timoshuk2024}.  Because the two order-parameter modes mix, the $\bm q$ dependence of an individual $|\Delta_\alpha|$ cannot be associated with only one of the two coherence lengths.  Therefore, FMP(cGL) cannot separately extract the two coherence lengths in the two-band case, and we compare FMP(eGL) with the real-space BdG method.

\begin{figure}[t]
\centering
\includegraphics[width=0.9\linewidth]{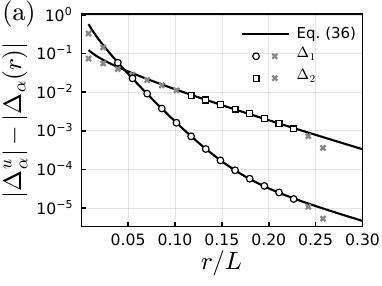}

\medskip

\includegraphics[width=0.9\linewidth]{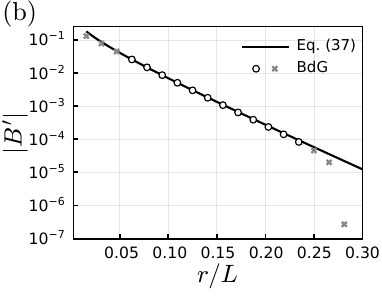}
\caption{(a) Fits to $|\Delta_1(r)|$ and $|\Delta_2(r)|$ and (b) fit to $B'(r)$ obtained with the real-space BdG method for the two-band model of Sec.~\ref{subsec:two-result} ($V_{11}=2.8$, $V_{22}=2.2$, $V_{12}=0.01$, $L=64$, $T/T_{\mathrm c}=0.644$).  The radial coordinate is normalized by $L=64$.  In panel (a), open circles for $\Delta_1$ and open squares for $\Delta_2$ indicate the data included in the fits.  In panel (b), open circles indicate the data included in the fits.  Gray crosses indicate the data excluded from the fits.}
\label{fig:bdg-fit-two}
\end{figure}

\begin{figure}[!t]
\centering
\includegraphics[width=0.9\linewidth]{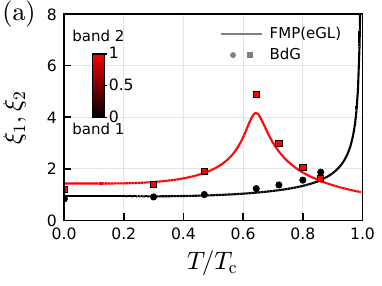}

\medskip

\includegraphics[width=0.9\linewidth]{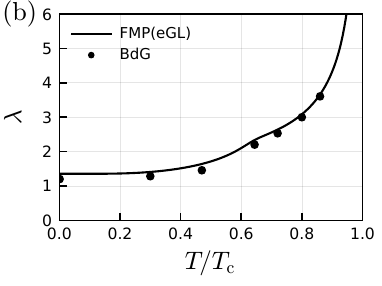}
\caption{Results for the two-band model with $V_{11}=2.8$, $V_{22}=2.2$, and $V_{12}=0.01$.  (a) Two coherence lengths obtained with FMP(eGL) from Eq.~\eqref{eq:xi-final}.  The color scale encodes the band character of each branch, obtained from the eigenvectors of the secular equation [Eq.~\eqref{eq:lin-eq}].  (b) Magnetic penetration depth obtained with FMP(eGL) from Eq.~\eqref{eq:lambda-final}.  Symbols show values obtained with the real-space BdG method.  In (a), the symbols are colored by the band weights obtained from $\cos\Theta$ and $\sin\Theta$ of Eq.~\eqref{eq:fit-two} using the same color scale.}
\label{fig:two-results}
\end{figure}

Representative fits to $|\Delta_1(r)|$, $|\Delta_2(r)|$, and $B'(r)$ obtained with the real-space BdG method [Eqs.~\eqref{eq:fit-two} and \eqref{eq:fit-B}] are shown in Fig.~\ref{fig:bdg-fit-two}.  Figure~\ref{fig:two-results} compares the two coherence lengths and the magnetic penetration depth obtained with FMP(eGL) and the real-space BdG method as functions of temperature.
The branch dominated by band 2, which has the weaker intraband pairing interaction, peaks near $T/T_{\mathrm c}\simeq0.64$.  This peak reflects the hidden critical point of the weaker band, i.e., the transition temperature at which band 2 would become superconducting without interband coupling, consistent with the hidden criticality reported for two-band BCS models \cite{Komendova2012}.

FMP(eGL) reproduces three features of the temperature dependence: the peak in the band-2-dominated branch, the slower growth of the band-1-dominated branch, and the monotonic increase of $\lambda$ toward $T_{\mathrm c}$.  The two methods agree well over the range shown for $\lambda$ and both coherence lengths.  The differences between the coherence lengths obtained with the two methods increase with $\xi$ because finite-size effects become more significant for larger coherence lengths (Appendix~\ref{app:finite-size}).  At the opposite limit, when a coherence length is only a few lattice constants, the lattice may be too coarse for the real-space BdG method to resolve the spatial variation of the order parameter, while neglected gradient terms may affect the values obtained with FMP(eGL).  Over a range of temperatures, $\lambda$ lies between the two coherence lengths, so the system is in a type-1.5 regime.  For example, at $T/T_{\mathrm c}=0.644$, FMP(eGL) gives $\xi_1=1.1$, $\lambda=2.3$, and $\xi_2=4.2$, which satisfy $\xi_1<\lambda<\xi_2$.

\section{Conclusion}
\label{sec:conclusion}

We have developed FMP(eGL), an extended GL method for calculating coherence lengths and the magnetic penetration depth in single-band and two-band models.  In this method, only the covariant-gradient terms are perturbatively treated, while the superconducting order parameters are fully taken into account.  For a two-band system, varying the free energy gives two coupled equations for $\psi_1$ and $\psi_2$.  Linearizing these equations around the uniform solution yields two coherence lengths.  The variation of the free energy with respect to the vector potential determines the magnetic penetration depth.

FMP(eGL) remains accurate at $T \ll T_{\mathrm c}$.  Its results agree well with those obtained with the real-space BdG method, whereas FMP(cGL), which uses relations derived from conventional GL theory, is reliable only when the order parameter is small.  Within the adopted covariant-gradient expansion, FMP(eGL) extracts the required coefficient functions from microscopic free energies at finite $\bm q$ and can therefore calculate these length scales when such free energies are available.  Because these free energies are evaluated on a momentum mesh, FMP(eGL) does not require a self-consistent calculation over the entire real-space lattice, unlike the real-space BdG method.  Its computational cost is comparable to that of FMP(cGL).  Like the real-space BdG method, FMP(eGL) is also applicable to multiband superconductors, whereas FMP(cGL) cannot calculate multiple coherence lengths in the multiband case.

\begin{acknowledgments}
We thank H. Matsunaga, R. Oiwa, and N. Witt for helpful discussions.
This work was supported by Grants-in-Aid for Scientific Research from JSPS (KAKENHI Grants No.~25H01252, No.~25H02115, No.~24K00581 and No.~24H00190), JST K-Program JPMJKP25Z7, and the RIKEN TRIP initiative (RIKEN Quantum, Advanced General Intelligence for Science Program, Many-body Electron Systems).
T.R.F. acknowledges financial support from the Program for Leading Graduate Schools (MERIT-WINGS).
\end{acknowledgments}

\appendix
\section{FMP(cGL): formulation within conventional GL theory}
\label{app:fmp}

\begin{figure}[!t]
\centering
\includegraphics[width=\columnwidth]{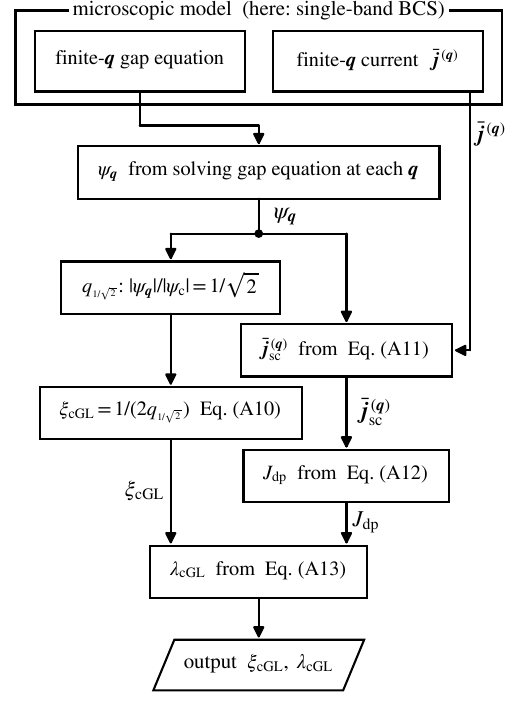}
\caption{Flowchart of FMP(cGL) in the single-band case.  Unlike FMP(eGL), FMP(cGL) determines the order parameter $\psi_{\bm q}$ self-consistently at each imposed momentum $\bm q$.}
\label{fig:flowchart-fmp}
\end{figure}

We summarize the FMP(cGL) formulas used for the single-band results in Sec.~\ref{subsec:single-result}, following the derivation in Ref.~\cite{Witt2024}.  Figure~\ref{fig:flowchart-fmp} provides an overview of the single-band FMP(cGL) procedure.  First, the order parameter $\psi_{\bm q}$ is determined self-consistently at each imposed momentum $\bm q$, and the coherence length $\xi_{\rm cGL}$ is extracted from the computed $|\psi_{\bm q}|$ using the $1/\sqrt{2}$ criterion.  We then obtain the depairing current density $J_{\rm dp}$ by maximizing the spatially averaged supercurrent with respect to $q$ and use $\xi_{\rm cGL}$ and $J_{\rm dp}$ to calculate the magnetic penetration depth $\lambda_{\rm cGL}$.

The equations used in this procedure are presented below.  We start from conventional GL theory.  For a single complex order parameter $\psi(\bm r)$, the GL free energy is
\begin{equation}
\begin{aligned}
\mathcal F_{\rm GL}=\int \dd\bm r\,
\Big[&F_n-\alpha |\psi|^2
 +\frac{b}{2}|\psi|^4
 +\frac{\hbar^2}{2m^*}|\D\psi|^2\\
&+\frac{(\nabla\times\bm A)^2}{2\mu_0}
\Big].
\end{aligned}
\label{eq:conventional-gl-free-energy}
\end{equation}
Here $\alpha$ and $b$ are the conventional GL coefficients and are positive below $T_{\mathrm c}$.  In this appendix, $\alpha$ does not denote a band index.  The quantity $m^*$ is the effective mass, and $\D$ is the covariant derivative defined in Sec.~\ref{subsec:single-formulation}.  The variation of $\mathcal F_{\rm GL}$ with respect to $\psi^*$ gives the conventional GL equation,
\begin{equation}
\left[-\frac{\hbar^2}{2m^*}\D^2-\alpha+b|\psi|^2\right]\psi=0.
\label{eq:single-gl-eq}
\end{equation}
For the uniform solution at $\bm A = \bm 0$ and $\bm q = \bm 0$, denoted by $\psi_{\mathrm c}$ as in the main text, Eq.~\eqref{eq:single-gl-eq} yields
\begin{equation}
|\psi_{\mathrm c}|^2=\frac{\alpha}{b} .
\label{eq:single-gl-uniform}
\end{equation}
For the uniform solution, the supercurrent is $\bm j=-(e^{*2}/m^*)|\psi_{\mathrm c}|^2\bm A$.  Combining this expression with Amp\`ere's law gives the London equation
\begin{equation}
\nabla^2\bm B=\frac{1}{\lambda_{\rm GL}^2}\bm B .
\label{eq:single-gl-london}
\end{equation}
The magnetic penetration depth is therefore
\begin{equation}
\lambda_{\rm GL}=\sqrt{
\frac{m^*b}{\mu_0 e^{*2}\alpha}
} .
\label{eq:single-gl-lambda}
\end{equation}

The conventional GL coherence length is obtained by writing $\psi=\psi_{\mathrm c}+\delta\psi$ and linearizing the GL equation. This gives
\begin{equation}
\left[2\alpha-\frac{\hbar^2}{2m^*}\nabla^2\right]\delta\psi=0 .
\label{eq:single-gl-linearized}
\end{equation}
Using the same convention as in the main text, we rewrite this equation as $\nabla^2\delta\psi=\xi_{\rm GL}^{-2}\delta\psi$.  This yields
\begin{equation}
\xi_{\rm GL}=\sqrt{\frac{\hbar^2}{4m^*\alpha}} .
\label{eq:single-gl-xi}
\end{equation}
This definition of $\xi_{\rm GL}$ differs by a factor of $\sqrt{2}$ from that obtained in the usual derivation of GL theory and is consistent with the procedure in the main text.
Equations~\eqref{eq:single-gl-lambda} and \eqref{eq:single-gl-xi} are the conventional single-band GL formulas.

FMP(cGL) uses a superconducting state in which the Cooper pairs have a finite center-of-mass momentum $\bm q$.
For a single order parameter, we write this state as
\begin{equation}
\psi_{\bm q}(\bm r)=|\psi_{\bm q}|
e^{\ii\bm q\cdot\bm r}.
\label{eq:ff-state}
\end{equation}
In contrast to the phase twist of Sec.~\ref{subsec:single-formulation}, in which the order-parameter magnitude is held fixed, here $|\psi_{\bm q}|$ is determined self-consistently for each $\bm q$.  Substituting Eq.~\eqref{eq:ff-state} into Eq.~\eqref{eq:conventional-gl-free-energy} at $\bm A=\bm 0$ gives the free-energy density
\begin{equation}
F_{\rm GL}(q)=F_n+
\left(-\alpha+\frac{\hbar^2q^2}{2m^*}\right)|\psi_{\bm q}|^2
+\frac{b}{2}|\psi_{\bm q}|^4 .
\label{eq:fmp-gl-free-energy}
\end{equation}
The self-consistent order parameter minimizes the free-energy density, $\psi_{\bm q}=\operatorname*{argmin}_{\psi}F_{\rm{GL}}(\bm q)$.  For the free-energy density of Eq.~\eqref{eq:fmp-gl-free-energy}, we obtain
\begin{equation}
|\psi_{\bm q}|^2=|\psi_{\mathrm c}|^2
\left(1-2\xi_{\rm cGL}^2q^2\right).
\label{eq:fmp-gap-q-square}
\end{equation}
Equation~\eqref{eq:fmp-gap-q-square} is derived from conventional GL theory and is therefore valid only when the order parameter is small.  This limitation explains the low-temperature deviations of the FMP(cGL) curves in Fig.~\ref{fig:single-results}.  Here $\xi_{\rm cGL}$ is the coherence length obtained with FMP(cGL).  Within conventional GL theory it coincides with $\xi_{\rm GL}$ of Eq.~\eqref{eq:single-gl-xi}, whereas in FMP(cGL) it is extracted directly from the computed $|\psi_{\bm q}|$.  Following Refs.~\cite{Witt2024,Kawamura2026}, we define $q_{1/\sqrt{2}}$ by $|\psi_{\bm q}|/|\psi_{\mathrm c}|=1/\sqrt{2}$.  Equation~\eqref{eq:fmp-gap-q-square} then gives $\xi_{\rm cGL}=1/(2q_{1/\sqrt{2}})$, and this $1/\sqrt{2}$ criterion is used for the FMP(cGL) results in Fig.~\ref{fig:single-results}.

For comparison, we determine $\xi_{\rm cGL}$ by fitting Eq.~\eqref{eq:fmp-gap-q-square} to the computed $|\psi_{\bm q}|$ at a few small momenta.  We also use the critical momentum $q_{\rm c}$, defined by $|\psi_{\bm q}|=0$, which gives $\xi_{\rm cGL}=1/(\sqrt{2}q_{\rm c})$.

FMP(cGL) determines the magnetic penetration depth from the current response \cite{Witt2024}.  We compute the spatially averaged supercurrent carried by Cooper pairs with center-of-mass momentum $\bm q$,
\begin{equation}
\bar{\bm j}_{\rm sc}^{(\bm q)}
=\bar{\bm j}^{(\bm q)}
-\left.\bar{\bm j}^{(\bm q)}\right|_{\psi=0}.
\label{eq:fmp-current}
\end{equation}
Within conventional GL theory, the current carried by the state in Eq.~\eqref{eq:ff-state} is $\bar j_{\rm sc}^{(q)}=(e^*\hbar/m^*)\,|\psi_{\bm q}|^2 \bm q$.  Inserting $|\psi_{\bm q}|^2$ from Eq.~\eqref{eq:fmp-gap-q-square} and maximizing with respect to $q$ gives the depairing momentum $q_{\rm dp}=1/(\sqrt{6}\,\xi_{\rm cGL})$ and the depairing current density
\begin{equation}
J_{\rm dp}=\max_q\,\bar j_{\rm sc}^{(q)}
=\frac{2}{3\sqrt{6}}\,\frac{|e^*|\hbar}{m^*}\,
\frac{|\psi_{\mathrm c}|^2}{\xi_{\rm cGL}} .
\label{eq:depairing-current}
\end{equation}
We eliminate $|\psi_{\mathrm c}|^2$ via Eqs.~\eqref{eq:single-gl-uniform} and \eqref{eq:single-gl-lambda} and use the flux quantum $\Phi_0=2\pi\hbar/|e^*|$ to obtain the following expression for the magnetic penetration depth in terms of the depairing current,
\begin{equation}
\lambda_{\rm cGL}=\sqrt{
\frac{\Phi_0}{3\sqrt{6}\,\pi\,\mu_0\,\xi_{\rm cGL}\,J_{\rm dp}}
} .
\label{eq:fmp-lambda-peak}
\end{equation}
We use Eq.~\eqref{eq:fmp-lambda-peak} to obtain the dashed FMP(cGL) curve for $\lambda$ in Fig.~\ref{fig:single-results}(b).  The magnetic penetration depth can also be obtained from the slope of the spatially averaged supercurrent with respect to $q$ at $q=0$, as in Ref.~\cite{Kawamura2026}.

\begin{figure}
\centering
\includegraphics[width=\columnwidth]{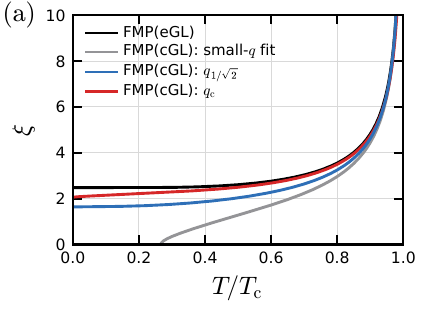}
\includegraphics[width=\columnwidth]{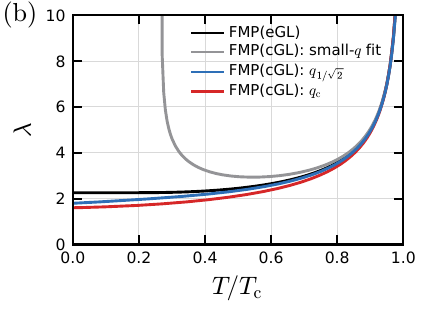}
\caption{Comparison of FMP(cGL) and FMP(eGL) for the single-band model.  (a) Coherence length and (b) magnetic penetration depth plotted against $T/T_{\mathrm c}$.  The FMP(cGL) results are obtained using the small-$q$ fit, the $1/\sqrt{2}$ criterion, and the $q_{\rm c}$ criterion.}
\label{fig:fmp-cgl-comparison}
\end{figure}

Figure~\ref{fig:fmp-cgl-comparison} compares $\xi$ and $\lambda$ obtained from the three FMP(cGL) procedures with those obtained from FMP(eGL).  For each FMP(cGL) procedure, $\lambda_{\rm cGL}$ is calculated from Eq.~\eqref{eq:fmp-lambda-peak} using the corresponding $\xi_{\rm cGL}$ and the same $J_{\rm dp}$.  At $T/T_{\mathrm c}=0.0005$, close to the zero-temperature limit, the small-$q$ fit breaks down.  The relative deviations of the $1/\sqrt{2}$ criterion and the $q_{\rm c}$ criterion from FMP(eGL) are 33.9\% and 16.9\% for $\xi$ and 20.4\% and 29.0\% for $\lambda$, respectively.

\begin{figure}
\centering
\includegraphics[width=\columnwidth]{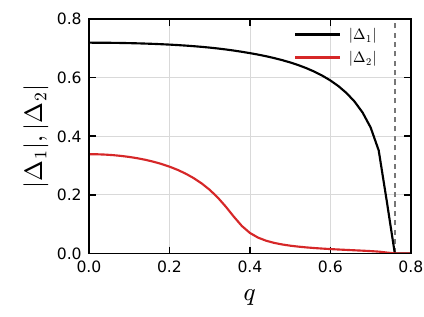}
\caption{Order-parameter magnitudes $|\Delta_1(q)|$ and $|\Delta_2(q)|$ plotted against $q$ for the two-band model of Sec.~\ref{subsec:two-result}.  The vertical dashed line indicates the position of $q_{\rm c}$.}
\label{fig:fmp-cgl-two-band}
\end{figure}

To illustrate why FMP(cGL) cannot separately determine the two coherence lengths in the two-band model, Fig.~\ref{fig:fmp-cgl-two-band} shows $|\Delta_1(q)|$ and $|\Delta_2(q)|$ plotted against $q$ for the model of Sec.~\ref{subsec:two-result}.
Unlike in the single-band case, the $q$ dependences of $|\Delta_1(q)|$ and $|\Delta_2(q)|$ do not follow the simple form of Eq.~\eqref{eq:fmp-gap-q-square}.  It is therefore not clear whether the small-$q$ fit or the $1/\sqrt{2}$ criterion is appropriate for these curves.  In addition, because the two order-parameter modes mix, a length scale extracted separately from each $|\Delta_\alpha(q)|$ cannot be directly assigned to one of the two coherence-length modes.  The two order parameters also vanish at the same critical momentum $q_{\rm c}$, so the $q_{\rm c}$ criterion yields only one coherence length.  Therefore, these FMP(cGL) procedures cannot separately determine the two coherence lengths in this two-band case.

\section{Finite-size effects on the extracted coherence lengths}
\label{app:finite-size}

To examine the system-size dependence of the length scales reported in Sec.~\ref{sec:results}, we perform the vortex calculation for the single-band model at $T/T_{\mathrm c}=0.4$ for $V=1.2$, $1.3$, $1.5$, $1.7$, $2.0$, $2.2$, $2.5$, and $2.8$ at $L=64$, $96$, and $128$.  We use the same fitting functions as in Sec.~\ref{subsec:realspace} and the same rule for choosing the fitting range for every $V$ and $L$.

As a guideline, the region within $r\simeq2\,\xi$ of the vortex center is excluded from the fit, where $\xi$ is the fitted coherence length.  The fitting range ends when the data no longer follow Eq.~\eqref{eq:fit-single}, before they turn upward due to finite-size effects.  As $L$ increases, a larger portion of the data can be included in the fit, so the outer edge of the fitting range does not decrease.
Table~\ref{tab:finite-size-check} summarizes the finite-size effects in terms of the ratio $\xi_L/\xi_{\rm eGL}$, where $\xi_L$ is the coherence length obtained by fitting the $L\times L$ data and $\xi_{\rm eGL}$ is the value obtained with FMP(eGL) at the same $V$.

For $V\ge1.7$, this ratio is approximately independent of $L$, so the coherence lengths extracted at $L=64$ are converged with respect to $L$.  For weaker couplings ($V\le1.5$), the ratio increases systematically with $L$ (from $0.50$ at $L=64$ to $0.75$ at $L=128$ for $V=1.2$, and from $0.78$ to $0.96$ for $V=1.5$).  This occurs because $|\Delta(r)|$ does not reach its bulk value in a finite system when the coherence length is comparable to or larger than the system size.  The ratio approaches $1$ as $L$ increases.

In contrast, at stronger couplings the ratio is independent of $L$ and reaches $1.23$ at $V=2.8$, where $\xi_{\rm eGL}$ is only a few lattice constants.  In this regime, the lattice is too coarse to resolve the spatial variation of the order parameter for the real-space BdG method, and neglected gradient terms may affect $\xi_{\rm eGL}$ obtained with FMP(eGL).  One or both of these effects may contribute to this difference.

\begin{table}[!htbp]
\caption{Finite-size effects on the coherence lengths obtained with the single-band real-space BdG method at $T/T_{\mathrm c}=0.4$.  Each value in this table is $\xi_L/\xi_{\rm eGL}$, where $\xi_{\rm eGL}$ is obtained with FMP(eGL).}
\label{tab:finite-size-check}
\begin{ruledtabular}
\begin{tabular}{cccc}
$V$ & $L=64$ & $L=96$ & $L=128$ \\
\hline
1.2 & $0.50$ & $0.64$ & $0.75$ \\
1.3 & $0.58$ & $0.72$ & $0.81$ \\
1.5 & $0.78$ & $0.90$ & $0.96$ \\
1.7 & $0.96$ & $1.02$ & $1.03$ \\
2.0 & $1.02$ & $1.02$ & $1.02$ \\
2.2 & $1.06$ & $1.07$ & $1.07$ \\
2.5 & $1.13$ & $1.14$ & $1.14$ \\
2.8 & $1.23$ & $1.23$ & $1.23$ \\
\end{tabular}
\end{ruledtabular}
\end{table}
\section{Nonlocal corrections to the extracted magnetic penetration depth}
\label{app:nonlocal}

FMP(eGL) determines the magnetic penetration depth from the London equation [Eq.~\eqref{eq:single-lambda} and its two-band counterpart, Eq.~\eqref{eq:lambda-final}].  The real-space BdG method instead determines $\lambda$ by fitting the magnetic field around a vortex, which shows the nonlocal current response.  The approximation that the electromagnetic response is local is valid when $\lambda$ is much larger than the range of the response kernel, which is set by $\xi$.  When $\kappa=\lambda/\xi$ is small, the nonlocal response becomes important, and the profile of $B(r)$ can differ from that predicted by the London equation \cite{Pippard1953}.

In the present calculations, the nonlocal response becomes important at small pairing strength $V$ and low $T/T_{\mathrm c}$.  In this regime, $B(r)$ changes sign at a finite distance from the vortex core.  This overscreening is characteristic of a nonlocal response \cite{SenarathYapa2019}.  The radius at which $B(r)$ changes sign is independent of the system size.  This shows that the sign change is not a finite-size effect.  Since $K_0(r/\lambda)$ is positive, the fit based on Eq.~\eqref{eq:fit-B} cannot extend beyond the range with negative $B(r)$.  As a result, only a small number of data points can be included in the fit.

We examine the accuracy of this restricted fit for the single-band model at $V=2.0$, $T/T_{\mathrm c}=0.4$, and $L=64$.  For these parameters, nonlocal effects are significant, while finite-size effects are negligible (Appendix~\ref{app:finite-size}).  The fit based on Eq.~\eqref{eq:fit-B} shown in Fig.~\ref{fig:single-results}(b) uses only nine data points.  As an independent check, we determine $\lambda$ from the relation between the azimuthally averaged supercurrent and vector potential, $j(r)$ and $a(r)$.

We use the gauge-invariant vector potential $\bm a=\bm A-(\Phi_0/2\pi)\nabla\varphi$, where $\varphi$ is the phase of the order parameter and $\Phi_0=2\pi$ is the flux quantum in the units used here.  We denote the azimuthal averages of the supercurrent and $\bm a$ by $j(r)$ and $a(r)$, respectively.  In the units of Sec.~\ref{subsec:realspace}, the effective two-dimensional form of the Pippard relation given by Eq.~(9) of Ref.~\cite{SenarathYapa2019} is
\begin{equation}
j(r) = -\frac{1}{2\pi\xi\mu_0\lambda^2}\int a(r')\,
\frac{\exp\!\left(-|\bm r-\bm r'|/\xi\right)}{|\bm r-\bm r'|}\,\dd^2r' ,
\label{eq:pippard-fit}
\end{equation}
where $|\bm r-\bm r'|$ is the in-plane distance between $\bm r$ and $\bm r'$.

We use $a(r)$ obtained with the real-space BdG method as input to Eq.~\eqref{eq:pippard-fit} and fit the calculated $j(r)$ to the data obtained with the real-space BdG method.  The fitting parameter is $\lambda$.  We fix $\xi$ to the value obtained with FMP(eGL) at the same temperature.  The fitting range begins outside the vortex core, as described in Sec.~\ref{subsec:cost-scaling}.  We exclude the outermost points because the averaged supercurrent becomes too small to resolve.  Because this fit uses $j(r)$ rather than $B(r)$, its outer edge is not set by the distance from the center at which $B(r)$ changes sign.

\begin{figure}
\includegraphics[width=\linewidth]{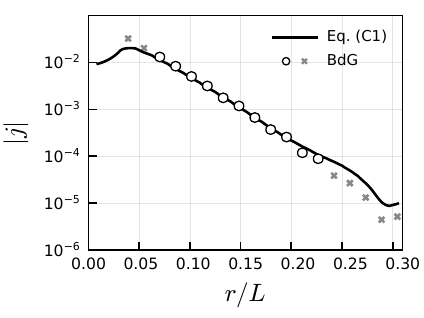}
\caption{Azimuthally averaged supercurrent $|j(r)|$ for the single-band model at $V=2.0$, $T/T_{\mathrm c}=0.4$, and $L=64$.  Open circles show the data included in the fit, and gray crosses show the excluded data.  The solid line is the fit obtained from Eq.~\eqref{eq:pippard-fit}.  We fix $\xi=1.79$, the value obtained with FMP(eGL), and obtain $\lambda=2.4$.  Lengths are in units of the lattice constant $a$.}
\label{fig:pippard-j}
\end{figure}

Figure~\ref{fig:pippard-j} shows that Eq.~\eqref{eq:pippard-fit} reproduces $j(r)$ over the full fitting range.  The fit gives $\lambda=2.4$.  This value differs by about $15\%$ from $\lambda=2.1$, obtained from the fit based on Eq.~\eqref{eq:fit-B} [Fig.~\ref{fig:single-results}(b)].  Therefore, the fit based on Eq.~\eqref{eq:fit-B} using only nine data points provides a reasonable estimate of the magnetic penetration depth even when nonlocal effects are strong.

The same check cannot be applied in the two-band case because the Pippard kernel contains more than one coherence length.  However, at low and intermediate temperatures, the value of $\lambda$ obtained from the fit based on Eq.~\eqref{eq:fit-B} agrees well with the value obtained with FMP(eGL) [Fig.~\ref{fig:two-results}(b)].  This agreement suggests that the fit based on Eq.~\eqref{eq:fit-B} has similar accuracy for the two-band system.

\bibliography{references}

\end{document}